\documentclass[showpacs,aps,prb,reprint,twocolumn]{revtex4-1}

\usepackage{commath}
\usepackage{amsmath}
\usepackage{graphicx}
\usepackage{hyperref}
\usepackage{txfonts}

\draft 

\begin{document}


\title{Possible instability of the Fermi sea against surface plasma oscillations}


\author{Hai-Yao Deng}
\email{h.deng@exeter.ac.uk}
\affiliation{School of Physics, University of Exeter, EX4 4QL Exeter, United Kingdom}

\begin{abstract} 
We derive a generic formalism for studying the energy conversion processes in bounded metals. Using this formalism we show that in the collision-less limit the Fermi sea of metals should experience an instability against surface plasma oscillations, which opens for the latter an intrinsic self-amplification channel. The origin of the instability is clarified as arising from novel effects resulting from the translation symetry breaking due to the very presence of surface. The amplification rate of this channel is analytically evaluated on the basis of energy conservation and the effects of losses are discussed. In particular, the unique role played by the surface in energy conversion is unveiled. In contrast with common wisdom and in line with observations, Landau damping is shown always overcompensated and therefore poses no serious issues in sub-wavelength plasmonics. 
\end{abstract}


\maketitle 

\section{introduction}
\label{section:1}
At low temperatures electrons reside in a sphere in the momentum space, known as the Fermi sea, provided they are free and independent~\cite{mermin1976}. Upon turning on their interactions, the Fermi sea can become unstable~\cite{pethick1988}. A familiar example is superconductivity, where even a tiny short-range attractive force between the electrons could destabilize the Fermi sea, resulting in an exponential growth of the Cooper pairing amplitude~\cite{philip2012}. Superconductivity represents a thermodynamic instability and shows up as a phase transition. Here we discuss an instability that occurs at a finite frequency and is manifested by an exponential while oscillatory increase of the amount of charges accumulated on the surfaces of metals. This instability is caused by surface plasma waves (SPWs) -- density undulations of electrons sustained by long-range Coulomb forces and propagating along metal surfaces.  

SPWs constitute a ubiquitous entity in metal optics~\cite{ebbesen2008,ozbay2006,zayats2005,barnes2003}. Systematical studies of SPWs begun over half a century ago when R. Ritchie in 1957 investigated the energy losses of electrons passing through metal foils~\cite{ritchie1957,ferrell1958}. A comprehensive understanding was soon accomplished of many fundamental properties of SPWs in the following decade or so~\cite{raether1988}. Since then studies on SPWs have become largely application oriented and remarkable progresses have been made in a plethora of areas in the past two decades~\cite{rothen1988,maier2007}. However,
most existing studies have presumed that the electrons underpinning SPWs move like a fluid within the hydrodynamic-Drude approach~\cite{pendry2015,pitarke2007,kik2007,fetter1986}. In this work we take a complementary perspective by assuming that the electrons move ballistically, for which the hydrodynamic description fails. This situation may also be of some practical interest. Many SPW-based applications are hampered by energy losses due to electronic collisions~\cite{khurgin2015}. It has been even argued that the Landau damping associated with excitation of electron-hole pairs alone would pose a sufficiently adverse factor in sub-wavelength plasmonics in deterring the operation of spaser - a plasmonic analogy of laser~\cite{khurgin2015}. In order to reduce such losses, materials of high quality are being actively sought~\cite{abajo2015,west2010,weick2017}. In these materials electrons could be ballistic to certain extent and a theory of SPWs underpinned with such ballistic electrons should then be of great value in guiding future experiments. Despite the interest, ballistic SPWs have so far received little attention.  

\begin{figure*}
\begin{center}
\includegraphics[width=0.95\textwidth]{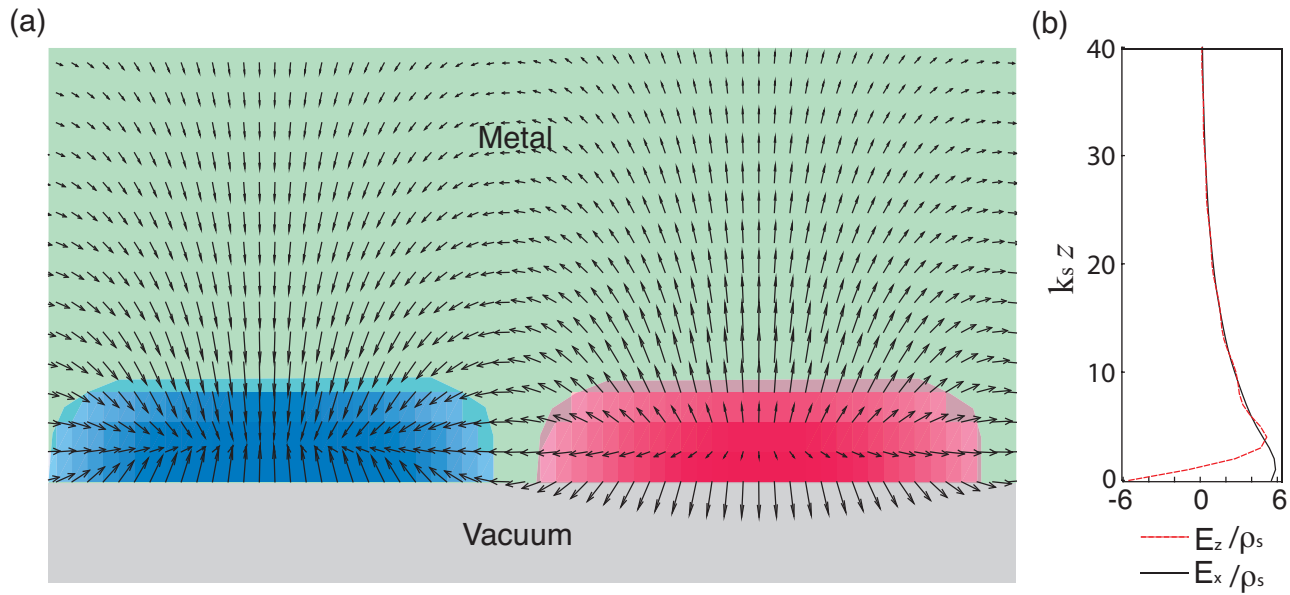}
\end{center}
\caption{Surface plasma waves (SPWs) supported on a semi-infinite metal (SIM) and the associated electric field. The SIM surface is located on the plane $z=0$. In (a), the arrows indicate the electric field generated by the charges which are indicated by the colors, of the SPWs. In (b), a plot of the electric field is displayed according to Eq.~(4) with $\rho_q \propto 1/(\omega^2_b - \omega^2_s)$, where $\omega_b$ is the bulk plasma wave dispersion relation and thus a function of $\sqrt{k^2+q^2}$ and $\omega_s$ is the SPW frequency. The exact form of $\rho_q$ is not necessary for the calculations performed in the present work. The apparent oscillation appearing in $E_z(z)$ is due to numerical inaccuracy. $k_s = \omega_s/v_F$, where $v_F$ denotes the Fermi velocity of the metal. \label{figure:f1}}
\end{figure*}

Recently~\cite{deng2016a} we studied ballistic SPWs in an ideal yet prototypical system, namely, a semi-infinite metal (SIM) occupying the half space $z\geq0$ with a geometric surface located at $z=0$, as shown in Fig.~\ref{figure:f1} (a). The metal was described by the jellium model~\cite{jellium} and inter-band transitions are accordingly neglected. By constructing an equation of motion for the charge density based on Boltzmann's theory, we discovered that SPWs in this system are unstable and can spontaneously amplify in the collision-less limit. The theory has also been extended to metal films and the same scenario occurs~\cite{deng2016b}. The amplification indicates a growing-up of the electrostatic potential energy of the system. Now that the total energy must be conserved, an increase of potential energy implies a decrease of the kinetic energy stored in the Fermi sea, thereby signifying an instability of the latter. This instability is obviously a consequence of the interplay between the long-range Coulomb interaction and translation symmetry breaking due to the surface. However, the equation of motion approach used in Ref.~\cite{deng2016a,deng2016b} does not allow us to penetrate directly into the physical mechanism by which the energy is actually transferred from the electrons to the waves. Such an energy conversion picture not only complements the equation of motion approach but also furnishes a physically transparent explanation of the instability and amplification scenario. 

The main purpose of the present work is to fill this gap of understanding. In this work, we provide a detailed picture of the energy conversion involved in ballistic SPWs supported on the surface of a SIM. We show that the instability is subsequent to the interplay between ballistic electronic motions and the surface. These motions allow SPWs to draw energy from the electrons when a surface is present. More specifically, we find that the electrical current density $\mathbf{J}(\mathbf{x},t)$ can be split in two disparate components, which we call $\mathbf{J}_{b}(\mathbf{x},t)$ and $\mathbf{J}_s(\mathbf{x},t)$, respectively. They are discriminated in many ways, for example by their correlations with the electric field $\mathbf{E}(\mathbf{x},t)$ generated by the charge density $\rho(\mathbf{x},t)$ of the system. It turns out that $\mathbf{J}_b(\mathbf{\mathbf{x},t})$ relates to $\mathbf{E}(\mathbf{x},t)$ in the same fassion as in a bulk system without surfaces, regardless of the value of the thermal electronic collision rate $\tau^{-1}$. For this reason, $\mathbf{J}_b(\mathbf{x},t)$ is called the bulk component, which would be partially captured in the hydrodynamic model but Landau damping. However, $\mathbf{J}_s(\mathbf{x},t)$ has no simple relation with $\mathbf{E}(\mathbf{x},t)$ and is totally absent from the hydrodynamic description. What is peculiar with $\mathbf{J}_s(\mathbf{x},t)$ is that, it would wholly disappear without the surface and thus represents genuine surface effects. We call it the surface component. We calculate the rate of growth of the electrostatic potential energy and equate it with the work done per unit time on the electrical currents by the electric field $\mathbf{E}(\mathbf{x},t)$ to obtain the self-amplification rate $\gamma_0$. This calculation is in spirit similar to Dawson's evaluation of the rate of Landau damping~\cite{dawson1962} but is more subtle due to the surface. We have then developed a generic formalism for the energy conversion processes involving surfaces. We find that, in spite of Landau damping, $\mathbf{J}_s(\mathbf{x},t)$ imparts a net amount of kinetic energy of the electrons to the waves and is responsible for the instability. The present formalism can also serve a general framework for discussing the losses due to inter-band transitions and radiation.

In the next section, we introduce a rigorous framework for studying the energy conversion in the presence of a surface. A generic equation of energy balance is established and employed to prove the instability of the Fermi sea of a SIM in later sections. The critical role of the surface in energy conversion, which has so far not been recognized, is unveiled and highlighted. In Sec.~\ref{section:3}, we prescribe the electronic distribution function, whose structure is analyzed in Sec.~\ref{section:4}. We split this function into a bulk component and a surface component, the definitions of which are quantitatively established. In Secs.~\ref{section:5} and \ref{section:6}, we evaluate the work done by the electric field on the electrons via the bulk and the surface components, respectively. It is shown that, the only effect of the bulk component is to bring about Landau damping; otherwise, no net transfer of energy would happen between the electrons and waves. This is so only for the presence of a surface. On the contrary, the surface component always imparts an amount of energy from the electrons to the waves and thus makes an intrinsic gain for SPWs. More interestingly, this gain always overcompensates for the Landau damping and only competes with the loss due to thermal electronic collisions. The intrinsic amplification rate is calculated in Sec.~\ref{section:7}. In Sec.~\ref{section:8}, we discuss the result and summarize the paper. Finally, an appendix is provided to illustrate some conceptual point.  

\section{Energy conversion with a surface}
\label{section:2}
The system to be studied is a SIM described in our previous work~\cite{deng2016a}, see Fig.~\ref{figure:f1} (a). In accord with the jellium model, we treat it as a free electron gas embedded in a static background of uniformly distributed positive charges and confined to the half space $z\geq 0$. In equilibrium it is neutral everywhere. Perturbing the system by for example a beam of light leads to a variation in the concentration of electrons and the appearance of a charge density $\rho(\mathbf{x},t)$. With no regard to the underlying dynamics of the charges, be it classical or quantum mechanical, the equation of continuity must hold, namely, $$(\partial_t+\tau^{-1})\rho(\mathbf{x},t) + \partial_{\mathbf{x}}\cdot \mathbf{j}(\mathbf{x},t) = 0,$$ where $\mathbf{j}(\mathbf{x},t)$ stands for the electrical current density solely due to the presence of an electric field $\mathbf{E}(\mathbf{x},t)$, $\tau$ denotes the relaxation time, which by definition approaches infinity in the collision-less limit. Throughout we write $\mathbf{x}=(x,y,z)$ and reserve $\mathbf{r} = (x,y)$ for planar coordinates. In the continuity equation, we have included a damping term $-\rho(\mathbf{x},t)/\tau$ to account for the microscopic thermal electrical currents that arise from electronic collisions that tend to equilibrate the system. These currents correspond to the collision integral in Boltzmann theory and have nothing to do with the macroscopic fields appearing in the equation~\cite{landau}. 

The effects of a surface are two-fold. Firstly, the surface scatters and redistributes the electrons, an aspect to be discoursed in the next section within Boltzmann-Fuchs formalism. Secondly, the surface prevents any electrons from escaping the metal and $\mathbf{j}(\mathbf{x},t)$ must identically vanish for $z<0$. Thus, we write $\mathbf{j}(\mathbf{x},t)=\Theta(z)\mathbf{J}(\mathbf{x},t)$, where $\Theta(z)$ denotes the Heaviside step function. With this prescription the equation of continuity becomes
\begin{equation}
(\partial_t+\tau^{-1})\rho(\mathbf{x},t) + \partial_{\mathbf{x}}\cdot \mathbf{J}(\mathbf{x},t) = -\delta(z)J_z(\mathbf{x}_0,t), \label{1}
\end{equation}
Here $\mathbf{x}_0 = (\mathbf{r},0)$ denotes a point on the surface, $\delta(z)$ is the Dirac function peaked on the surface and $J_z(\mathbf{x},t)$ denotes the z-component of $\mathbf{J}(\mathbf{x},t)$. Equation (\ref{1}) can also be derived other ways (see Appendix \ref{sec:b}) and serve as the equation of motion for $\rho(\mathbf{x},t)$ if we express $\mathbf{J}(\mathbf{x},t)$ as a functional of $\rho(\mathbf{x},t)$ by means of Maxwell's equations. As discussed in Ref.~\cite{deng2016a,deng2016b}, bulk plasma waves, for which $J_z(\mathbf{x}_0,t)=0$, are governed only by the left hand side of this equation, while SPWs are described by solutions with non-vanishing $J_z(\mathbf{x}_0,t)$. If $J_z(\mathbf{x}_0,t)$ identically vanishes, the surface will be completely severed from the rest of the metal. In this case, by whatever external stimuli, e.g. a grazing charged particle, no charges can build up on the surface and SPWs can not be excited.  

We prescribe all field quantities in the form of a plane wave propagating along positive $x$ direction with a complex frequency $\omega = \omega_s+i\gamma$.  We write $\rho(\mathbf{x},t)=\left[\rho(z)e^{i(kx-\omega t)}\right]'$, $\mathbf{J}(\mathbf{x},t)=\left[\mathbf{J}(z)e^{i(kx-\omega t)}\right]'$ and $\mathbf{E}(\mathbf{x},t)=\left[\mathbf{E}(z)e^{i(kx-\omega t)}\right]'$, where $k\geq0$ denotes the wave number and a prime takes the real part of a quantity. As $\rho(z)$ exists only in half of the space, we also introduce a cosine Fourier transform like this, $$\rho_q = \int^{\infty}_0 dz~\cos(qz)\rho(z).$$ For SPWs, $\rho_q$ may be taken real-valued~\cite{deng2016b} and it only weakly depends on $q$ for not so large $q$. We then put $\rho_q \approx \rho_s$, where $\rho_s$ may be called the surface charge density. A cut-off $q_c$ has to be imposed on $q$ to reflect on the fact that $\rho(z)$ can not vary significantly over the mean inter-particle spacing $\sim n^{-1/3}$; otherwise, the jellium model would break down. Thus, we take $\rho_{q>q_c} \approx 0$ and $q_c\sim n^{1/3}$.~\cite{note2} Here $n$ denotes the mean concentration of electrons. With this prescription, $\rho(z)$ spreads over a layer of thickness of the order of $1/q_c$ within the surface, as required in the complete theory~\cite{deng2016a} and seen in Fig~\ref{figure:f1}. In terms of the characteristic plasma frequency of the metal, $\omega_p = \sqrt{4\pi ne^2/m}$, with $e$ being the charge and $m$ the mass of an electron, we have $\omega_s\approx\omega_p/\sqrt{2}$ as an approximation. We can calculate $\gamma$ by the principle of energy balance.

The electrostatic potential energy of the system is given by $$E_p(t) = \frac{1}{2}\int d^3\mathbf{x}~\rho(\mathbf{x},t)\phi(\mathbf{x},t),$$ where $\phi(\mathbf{x},t)$ is the electrostatic potential satisfying $$\partial^2_{\mathbf{x}}\phi(\mathbf{x},t)+4\pi\rho(\mathbf{x},t)=0.$$ One may be tempted to think that the rate of change of $E_p(t)$ can be directly calculated as the negative of the work done per unit time by the electric field $\mathbf{E}(\mathbf{x},t) = -\partial_{\mathbf{x}}\phi(\mathbf{x},t)$ on the electrons, which is, however, not true. This is because $E_p(t)$ does not count all the potential energy in the system. Specifically, it does not include the surface potential energy, which may be written $$E_s(t) = \int d^3\mathbf{x}~\rho(\mathbf{x},t)\phi_s(\mathbf{x}),$$ where $\phi_s(\mathbf{x})$ denotes the surface potential. For an ideal surface, $\phi_s(\mathbf{x})$ should vanish in the metal but rise to infinity everywhere on the surface so that no electrons can escape the metal. In the electrostatic and collision-less limit, energy conservation dictates that $$\dot{E}_p(t) = -\dot{E}_s(t)-\dot{E}_k(t) = \int d^3\mathbf{x}~\mathbf{J}(\mathbf{x},t)\cdot\mathbf{E}_s(\mathbf{x}) - P_b(t),$$ where the over-dot takes the time derivative, and $\mathbf{E}_s(\mathbf{x})=-\partial_{\mathbf{x}}\phi_s(\mathbf{x})$ as well as $$P_b(t) = \int d^3\mathbf{x}~\mathbf{J}(\mathbf{x},t)\cdot\mathbf{E}(\mathbf{x},t).$$ This relation explains why $\dot{E}_p$ is not given by $-P_b(t)$. The details of $\phi_s(\mathbf{x})$ and $\mathbf{E}_s(\mathbf{x})$, however, can hardly be known and could vary greatly from one sample to another. 

Notwithstanding, since the surface effects have been totally incorporated in the continuity equation, we can deduce a complete equation of energy balance from it. To this end, we multiply Eq.~(\ref{1}) by $\phi(\mathbf{x},t)$ and integrate it over $\mathbf{x}$. As both $\rho(\mathbf{x},t)$ and $\phi(\mathbf{x},t)$ evolve by the factor $e^{i(kx-\omega t)}$, we have $$\int d^3\mathbf{x}~\phi(\mathbf{x},t)\partial_t\rho(\mathbf{x},t)=\dot{E}_p(t).$$ Further, by integration by parts, $$\int d^3\mathbf{x}~\phi(\mathbf{x},t)\partial_{\mathbf{x}}\cdot \mathbf{J}(\mathbf{x},t)=P_b(t).$$ Similarly, from the right hand side of Eq.~(\ref{1}) it arises $$P_s(t) = \int d^3\mathbf{x}~\phi(\mathbf{x},t)\delta(z)J_z(\mathbf{x}_0,t) $$ which signifies genuine surface effects. As far as we are concerned, this term has not been noticed in existing work. It can be rewritten 
$$P_s(t) =  \int d^2\mathbf{r}~J_z(\mathbf{x}_0,t)\phi(\mathbf{x}_0,t).$$ Combining the above expressions, we arrive at 
\begin{equation}
\left(\frac{2}{\tau}+\partial_t\right)E_p(t) = - P_b(t) -P_s(t), \label{2}
\end{equation}
which is the energy balance equation of the system. 

Let us write the areal density of $E_p(t)$ and $P_{b,s}(t)$ as $\mathcal{E}_p(t)$ and $\mathcal{P}_{b,s}(t)$, respectively. We can perform the integration over $\mathbf{r}$ to get
\begin{eqnarray}
\mathcal{P}_{b,s}(t) = \frac{e^{2\gamma t}}{2} \mathcal{P}_{b,s}, \quad \mathcal{E}_p(t) = \frac{e^{2\gamma t}}{2} \mathcal{E}_p,
\end{eqnarray}
where $\mathcal{P}_{b,s}$ are given by
\begin{eqnarray}
\mathcal{P}_b &=& \int dz~\left[\mathbf{J}'(z)\cdot \mathbf{E}'(z)+ \mathbf{J}''(z)\cdot \mathbf{E}''(z)\right], \\
\mathcal{P}_s &=& J'_z(0) \phi'(0)+J''_z(0) \phi''(0).
\end{eqnarray}
Here $\mathbf{J}'(z)$ denotes the real part of $\mathbf{J}(z)$ and $\mathbf{J}''(z)$ the imaginary part, similarly for $\mathbf{E}(z)$ and other quantities. Analogously, we obtain
\begin{equation}
\mathcal{E}_p = \frac{1}{2} \int dz ~ \left[\rho'(z) \phi'(z)+\rho''(z) \phi''(z)\right].
\end{equation}
Taking $\rho(z)$ to be real, the terms involving the imaginary parts are then all gone. We thus obtain
\begin{eqnarray}
&~&\mathcal{P}_b = \int dz~\left[\mathbf{J}'(z)\cdot \mathbf{E}'(z)+ \mathbf{J}''(z)\cdot \mathbf{E}''(z)\right], \nonumber \\&~& \mathcal{P}_s = J'_z(0) \phi(0), \quad \mathcal{E}_p = \frac{1}{2} \int dz ~ \rho(z) \phi(z). 
\end{eqnarray}
Equation~(\ref{2}) can now be transformed in the following form
\begin{eqnarray}
2\gamma_0  = - (\mathcal{P}_b + \mathcal{P}_s)/\mathcal{E}_p, \quad \gamma_0 = \left(\frac{1}{\tau}+\gamma\right). \label{key}
\end{eqnarray} 
This is a key equation of the present paper. We shall show that $\gamma_0$ is always non-negative for surface plasma waves. 

By the laws of electrostatics, we find the potential given by $$ \phi(z) = \frac{2\pi}{k} \int dz'e^{-k\abs{z-z'}} \rho(z'),$$ from which it follows that $$\phi(0) = \frac{2\pi}{k} \xi, \quad \xi = \int^\infty_0 dz ~ e^{-kz}\rho(z).$$ As for the electric field, we write it as $\mathbf{E}(z)=\left(-iE_x(z),E_z(z)\right).$ In terms of $\rho_q$, we have 
\begin{equation}
\begin{pmatrix}
E_x(z)\\
E_z(z)
\end{pmatrix}
=\int^{\infty}_0 dq~\frac{4k\rho_q}{k^2+q^2} ~ 
\begin{pmatrix}
2\cos(qz)- e^{-kz} \\
2(q/k)\sin(qz) - e^{-kz}
\end{pmatrix}. \label{4}
\end{equation}
Thus, $\mathbf{E}'(z) = \left(0,E_z(z)\right)$ and $\mathbf{E}''(z) = \left(-E_x(z),0\right)$. With this we can rewrite 
\begin{equation}
\mathcal{P}_b = \int dz ~ \left[J'_z(z)E_z(z) - J''_x(z)E_x(z)\right].
\end{equation}

Generally the charge density $\rho(z)$ spreads over a layer a few multiples of $v_F/\omega_p = k^{-1}_p$ thick within the surface; see the example displayed in Fig.~\ref{figure:f1} (a). One can show that $E_x(z)\approx E_z(z)\approx E(z)=2\pi\rho_se^{-kz}$ outside the layer, while in the layer $E_x(z)$ and $E_z(z)$ are distinctly different and take opposite signs. Nonetheless, this layer makes a negligible contribution, of the order of $\kappa = k/k_p$, to the electrostatic potential energy $E_p$. This point becomes clear if we write $E_p(t) = \frac{1}{8\pi}\int d^3\mathbf{x}~\mathbf{E}^2(\mathbf{x},t)$. As an approximation, we may neglect the contribution of this layer and obtain 
\begin{equation}
\mathcal{E}_p \approx\frac{\pi\rho^2_s}{k},
\end{equation} 
Upon substituting this expression in Eq.~(\ref{key}), we end up with an equation for $\gamma_0$, since $\mathcal{P}_{b,s}$ are functions of $\gamma_0$. In order to find out $\gamma_0$, what remains to be done is to work out $\mathbf{J}(\mathbf{x},t)$ and use it to calculate $\mathcal{P}_{b,s}$.  

As an illustration of Eq.~(\ref{key}), let us apply it to the Drude model, by which $\rho(z) = \rho_s\delta(z)$ and $\mathbf{J}(z) = (i/\bar{\omega})(\omega^2_p/4\pi) \mathbf{E}(z)$, where $\bar{\omega} = \omega_s+i\gamma_0$ and $\mathbf{E}(z) \approx (-i,1)E(z)$ outside the surface layer. Assuming $\gamma_0/\omega_s\ll 1$ and then $i/\bar{\omega} \approx \gamma_0/\omega^2_s+i/\omega_s$, we find $\mathbf{J}(z) = \frac{\omega^2_p/\omega^2_s}{4\pi}\left(\omega_s-i\gamma_0,i\omega_s+\gamma_0\right)E(z)$. With this we find $\mathcal{P}_b = \frac{\gamma_0}{2\pi}\frac{\omega^2_p}{\omega^2_s} \int^\infty_0 dz~E^2(z) = \gamma_0 \frac{\omega^2_p}{\omega^2_s}\mathcal{E}_p$, where $\mathcal{E}_p = \rho_s\phi(0)/2$. Similarly, we find $\mathcal{P}_s = - \gamma_0 \frac{\omega^2_p}{\omega^2_s}\mathcal{E}_p$. Thus, $\mathcal{P}_b+\mathcal{P}_s = 0$ and $\gamma_0 = 0$ for the Drude model, agreeing with the equation of motion approach for the same model. Even for this simple model, the conventional picture of SPWs is incorrect. According to this picture, one would wrongly assume that the SPW damping is due to energy transfer between the electrons and the waves, by way of $\mathcal{P}_{b,b}$. The present calculation, however, shows that there is no net transfer of energy and the damping is solely caused by the presence of thermal currents that drives the system toward thermodynamic equilibrium. 

\section{The Electronic Distribution Function}
\label{section:3}
We ignore inter-band transitions and use Boltzmann's theory to study the electrical responses of the system. Surface scatters electrons. In principle, such scattering can be handled with a microscopic surface potential $\phi_s(\mathbf{x})$; see Appendix. However, $\phi_s(\mathbf{x})$ varies from one sample to another and is rarely known in practice. Alternatively, those effects may be dealt with using phenomenological boundary conditions.~\cite{fuchs1938,sondheimer1948,ziman1960,kaganov1997,jones2003} This is possible because $\phi_s(\mathbf{x})$ acts only on the surface and in the bulk the electronic distribution function $f(\mathbf{x},\mathbf{v},t)$ obtained as solutions to Boltzmann's equation can be written down without explicitly referring to the surface. A few parameters shall occur in the solutions and their values reflect on surface scattering. In the present paper, we will follow this approach to study the electrical responses of a SIM. 

As usual we divide the distribution function in two terms, $$f(\mathbf{x},\mathbf{v},t)=f_0\left(\varepsilon(\mathbf{v})\right) + g(\mathbf{x},\mathbf{v},t),$$ where $\varepsilon(\mathbf{v})=\frac{m}{2}\mathbf{v}^2$ is the kinetic energy of an electron while $f_0(\varepsilon)$ is the Fermi-Dirac function giving the equilibrium distribution and $g(\mathbf{x},\mathbf{v},t)$ denotes the deviation. Let us write $g(\mathbf{x},\mathbf{v},t)=$~Re$\left[g(\mathbf{v},z)e^{i(kx-\omega t)}\right]$. In the regime of linear response, Boltzmann's equation reads
\begin{equation}
\frac{\partial g(\mathbf{v},z)}{\partial z} + \lambda^{-1}~g(\mathbf{v},z) +
 ef'_0(\varepsilon)\frac{\mathbf{v}\cdot\mathbf{E}(z)}{v_z} = 0, 
\label{6}
\end{equation} 
where $\lambda = iv_z/\tilde{\omega}$ with $\tilde{\omega} = \bar{\omega} - kv_x$ and $\bar{\omega} = \omega + i/\tau$, and $f'_0(\varepsilon) = \partial_\varepsilon f_0(\varepsilon)$. The general solution is given by 
\begin{equation}
g(\mathbf{v},z) = e^{-\frac{z}{\lambda}}\left(C(\mathbf{v})-\frac{ef'_0\mathbf{v}}{v_z}\cdot \int^z_0~dz'~e^{\frac{z'}{\lambda}}~\mathbf{E}(z')\right), \label{7}
\end{equation}
where $C(\mathbf{v}) = g(\mathbf{v},0)$ is the non-equilibrium deviation on the surface to be determined by boundary conditions. We require $g(\mathbf{v},z)=0$ distant from the surface, i.e. $z\rightarrow\infty$. For electrons moving away from the surface, $v_z>0$, this condition is automatically fulfilled. For electrons moving toward the surface, $v_z<0$, it leads to
\begin{equation}
C(\mathbf{v}) =\frac{ef'_0\mathbf{v}}{v_z}\cdot \int^{\infty}_0~dz'~e^{z'/\lambda}\mathbf{E}(z'), \quad v_z<0, 
\end{equation}
yielding
\begin{equation}
g(\mathbf{v},z) = \frac{ef'_0\mathbf{v}}{v_z}\cdot \int^{\infty}_zdz' ~e^{\frac{z'-z}{\lambda}}~\mathbf{E}(z'), \quad v_z<0. 
\end{equation}
To determine $C(\mathbf{v})$ for $v_z>0$, the boundary condition at $z=0$ has to be used, which, whoever, depends on surface properties. We adopt a simple picture first conceived by Fuchs~\cite{fuchs1938}, according to which a fraction $p$ (Fuchs parameter varying between zero and unity) of the electrons impinging on the surface are specularly reflected back, i.e. 
\begin{equation}
g(\mathbf{v},z=0)=p~g(\mathbf{v}_-,z=0), \quad \mathbf{v}_- = (v_x,v_y,-v_z), \quad v_z\geq 0. \label{bc} 
\end{equation}
Note that this condition is identical with the condition used in Ref.~\cite{deng2016a,deng2016b} at $p=0$ but differs otherwise. It follows that 
\begin{equation}
C(\mathbf{v}) = - p~\frac{ef'_0\mathbf{v}_-}{v_z}\cdot \int^{\infty}_0dz' ~e^{-\frac{z'}{\lambda}}~\mathbf{E}(z'), \quad v_z\geq0. \label{la}
\end{equation} 
Equations (\ref{7}) - (\ref{la}) fully specify the distribution function for the electrons. The electrical current density is calculated as
\begin{equation}
\mathbf{J}(z) = (m/2\pi\hbar)^3\int d^3\mathbf{v} ~e\mathbf{v}g(\mathbf{v},z). \label{J} 
\end{equation}
It should be pointed out that the charge density is not given by $$\tilde{\rho}(\mathbf{x},t)=(m/2\pi\hbar)^3~\mbox{Re}\left[e^{i(kx-\omega t)}\int d^3\mathbf{v}~eg(\mathbf{v},z)\right],$$ which differs from the actual density $\rho(\mathbf{x},t)$ by a part localized on the surface. Actually, $\mathbf{J}(\mathbf{x},t)$ and $\tilde{\rho}(\mathbf{x},t)$ obey the equation $$(\partial_t+1/\tau)\tilde{\rho}(\mathbf{x},t)+\partial_{\mathbf{x}}\cdot\mathbf{J}(\mathbf{x},t)=0,$$ of which no SPWs are admitted, rather than the equation of continuity [c.f. Eq.~(\ref{1})]. 

\section{Decomposition into Bulk and Surface Components and the positiveness of $\gamma_0$}
\label{section:4}
The distribution function provided in Eqs.~(\ref{7}) - (\ref{la}) possesses a notable structure, which we reveal in this section. To this end, we substitute the expression of $\mathbf{E}(z)$ given by Eq.~(\ref{4}) into (\ref{7}) - (\ref{la}) and perform the integration over $z'$. We find that $g(\mathbf{v},z)$ can be split into two parts, one denoted by $g_b(\mathbf{v},z)$ while the other by $g_s(\mathbf{v},z)$. They are given by
\begin{eqnarray}
g_b(\mathbf{v},z) &=& -ef'_0\int^\infty_0 dq \frac{4\rho_q}{k^2+q^2}\times \nonumber \\&~& \quad \left(2F_+\cos(qz)+2iF_-\sin(qz)-F_0e^{-kz}\right), \label{gb}
\end{eqnarray}
where we have introduced the following functions,
\begin{equation}
F_\pm(\mathbf{K},\bar{\omega},\mathbf{v}) = \frac{1}{2}\left[\frac{\mathbf{K}\cdot\mathbf{v}}{\bar{\omega} - \mathbf{K}\cdot\mathbf{v}}\pm \frac{\mathbf{K}\cdot\mathbf{v}_-}{\bar{\omega} - \mathbf{K}\cdot\mathbf{v}_-}\right], \quad \mathbf{K} = (k,0,q).
\end{equation}
Note that $F_{+/-}$ is an even/odd function of $v_z$. In addition, 
\begin{equation}
F_0(\mathbf{k},\bar{\omega},\mathbf{v}) = \frac{\mathbf{k}^*\cdot\mathbf{v}}{\bar{\omega} - \mathbf{k}^*\cdot\mathbf{v}}, \quad \mathbf{k}^* = (k,0,ik). 
\end{equation}
Moreover, we have
\begin{eqnarray}
&~&g_s(\mathbf{v},z) = \Theta(v_z)(-ef'_0)e^{i\frac{\bar{\omega}z}{v_z}}\int^\infty_0 dq\frac{4\rho_q}{k^2+q^2}\times \label{gs} \\ &~& \quad \left[F_0(\mathbf{k},\bar{\omega},\mathbf{v})-pF_0(\mathbf{k},\bar{\omega},\mathbf{v}_-) + 2(p-1)F_+(\mathbf{K},\bar{\omega},\mathbf{v})\right]. \nonumber
\end{eqnarray}
At this stage, it is clear that $\gamma_0\geq 0$; otherwise, the factor $\exp(i\tilde{\omega}z/v_z) \propto \exp(-\gamma_0z/v_z)$  contained in $g_s(\mathbf{v},z)$ would diverge far away from the surface for any electrons departing the surface. This is nothing but a consequence of the causality principle. We shall confirm this by direct calculation of the energy conversion. 

Hereafter we call $g_b(\mathbf{v},z)$ the bulk component and $g_s(\mathbf{v},z)$ the surface component, for their disparate dependences on the presence of surface. For a bulk metal without the surface, i.e. if we send the surface to $-\infty$ or in other words replace $z$ by $z+z_0$ with $z_0\rightarrow\infty$, $g_s(\mathbf{v},z)$ will disappear identically whereas $g_b(\mathbf{v},z)$ does not, meaning that $g_s(\mathbf{v},z)$ signifies genuine surface effects while $g_b(\mathbf{v},z)$ gives the electrical responses of a bulk system. Actually, $g_b(\mathbf{v},z)$ has exactly the same form as the distribution function for a bulk metal in the presence of an electric field given in Eq.~(\ref{4}). One might think that $g_s(\mathbf{v},z)$ is insignificant. However, to the contrary it is indispensable in ensuring the boundary conditions (\ref{bc}) and hence constitutes an integral part of the complete electrical responses of a bounded system. 

We note that only electrons departing the surface contribute to $g_s(\mathbf{v},z)$. Those electrons either thermally emerge from the surface or have been bounced back (with probability $p$). We also note that $g_s(\mathbf{v},z)$ possess a phase factor $e^{i\varphi(z)}$, where $\varphi(z) = \omega_s z/v_z$. Physically, this phase is acquired when an electron leaves the surface and travels to the depth $z$ without suffering a collision. It is exactly these ballistic motions that are totally beyond the scope of the hydrodynamic-Drude model. 

Accordingly the current density $\mathbf{J}(z)$ also splits into a bulk and surface part, denoted by $\mathbf{J}_b(z)$ and $\mathbf{J}_s(z)$ respectively. They are defined via Eq.~(\ref{J}) with $g(\mathbf{v},z)$ replaced by $g_{b/s}(\mathbf{v},z)$. If we discard $\mathbf{J}_s(z)$, expand $F_\pm(\mathbf{K},\bar{\omega},\mathbf{v})$ in a series of $\mathbf{K}\cdot\mathbf{v}/\bar{\omega}$ and similarly $F_0(\mathbf{k},\bar{\omega},\mathbf{v})$ in $\mathbf{k}^*\cdot\mathbf{v}/\bar{\omega}$, and retain only the first terms in the series, we will then recover the current density expected of the hydrodynamic-Drude model. As $\mathbf{J}_s(z)$ is essential in gratifying the boundary condition (\ref{bc}), the hydrodynamic model is inadequate. In the next two sections, we calculate their contributions to $\mathcal{P}_{b/s}$ and show that, while $\mathbf{J}_b(z)$ produces what is expected of the hydrodynamic-Drude model apart from Landau damping, $\mathbf{J}_s(z)$ warrants a positive $\gamma_0$ and leads to an incipient instability of the system. 

\begin{figure*}
\begin{center}
\includegraphics[width=0.95\textwidth]{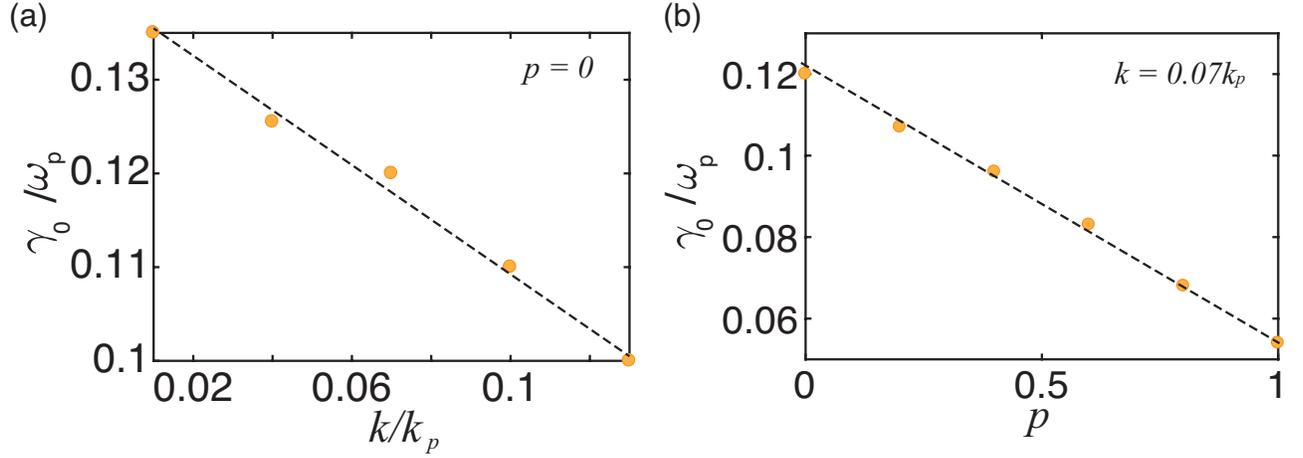}
\end{center}
\caption{The rate of gain $\gamma_0/\omega_p$ belonging with the intrinsic channel plotted against (a) the wavenumber $k/k_p$ and (b) the \textit{Fuchs} parameter $p$. Here $k_p = \omega_p/v_F$. Dots: $\gamma_0$ obtained by numerically solving Eq.~(\ref{key2}). Dashed line: a linear fitting as given in Sec. VII while also serves as a guide for the eye. \label{figure:f2}}
\end{figure*} 

\section{Electrical Work via $\mathbf{J}_b$}
\label{section:5}
Our purpose here is to calculate the contribution of $\mathbf{J}_b(z)$ to $\mathcal{P}_{b/s}$ and show that this contribution would vanish if Landau damping was excluded. Let us denote the contribution by $\mathcal{P}_{,b} = \mathcal{P}_{b,b}+\mathcal{P}_{s,b}$, where $\mathcal{P}_{s,b} = J'_{b,z}(0) \phi(0)$ and 
\begin{eqnarray}
\mathcal{P}_{b,b} = \int dz ~ \left[J'_{b,z}(z)E_z(z) - J''_{b,x}(z)E_x(z)\right],
\end{eqnarray}
Using Eq.~(\ref{gb}) and the parity of $F_{\pm}$ with respect to $v_z$, we obtain
\begin{equation}
J_{b,x}(z) = \int \mathcal{D}q\mathcal{D}^3\mathbf{v} ~ v_x\left[2F_+\cos(qz)-F_0e^{-kz}\right]. \label{jbx}
\end{equation}
where we have defined $$\int \mathcal{D}q\mathcal{D}^3\mathbf{v} ... = \left(\frac{m}{2\pi\hbar}\right)^3\int^{\infty}_0dq \frac{4\rho_q}{k^2+q^2}\int d^3\mathbf{v}\left(-e^2f'_0\right) ...$$ as a shortcut. Similarly,
\begin{equation}
J_{b,z}(z) = \int \mathcal{D}q\mathcal{D}^3\mathbf{v} ~ v_z\left[2iF_-\sin(qz)-F_0e^{-kz}\right]. \label{jbz}
\end{equation}
By expansion of $F_0$ in a series of $\mathbf{k}^*\cdot\mathbf{v}/\bar{\omega}$ and assuming that $\bar{\omega}$ be real, one can show that $\int d^3\mathbf{v} f'_0 v_x F_0$ is real whereas $\int d^3\mathbf{v} f'_0 v_z F_0$ is imaginary. With this, it follows from Eqs.~(\ref{jbx}) and (\ref{jbz}) that $J''_{b,x}(z)$ and $J'_{b,z}(z)$ would vanish if $\bar{\omega}$ were real. As such, $\mathcal{P}_{b/s,b}$ would also vanish under this assumption. 

However, $\bar{\omega}$ is not real but with an imaginary part $\gamma_0$. To calculate $\mathcal{P}_{b/s,b}$ under this circumstance, we find it instructive to rewrite $F_{\pm} = F^\text{D}_{\pm} + F^\text{L}_{\pm}$, with $F^\text{D}_+ = \frac{kv_x}{\bar{\omega}},~ F^\text{D}_- = \frac{qv_z}{\bar{\omega}},$ and 
\begin{equation}
F^\text{L}_\pm = \frac{1}{2}\left[\frac{(\mathbf{K}\cdot\mathbf{v})^2}{1-\mathbf{K}\cdot\mathbf{v}/\bar{\omega}}\pm \frac{(\mathbf{K}\cdot\mathbf{v}_-)^2}{1-\mathbf{K}\cdot\mathbf{v}_-/\bar{\omega}}\right]. \label{fl}
\end{equation}
Further taking $F_0 \approx \mathbf{k}^*\cdot\mathbf{v}/\bar{\omega}$, which is valid at long wavelengths $kv_F/\omega_p < 1$, we can split $\mathbf{J}_b(z)$ into a Drude term and an extra term as follows,
\begin{equation}
\mathbf{J}_b(z) = \frac{i}{\bar{\omega}}\frac{\omega^2_p}{4\pi} \mathbf{E}(z) + \mathbf{J}_L(z),
\end{equation}
where $\mathbf{J}_L(z)$ is given by
\begin{eqnarray}
J_{L,x}(z) &=& \int \mathcal{D}q\mathcal{D}^3\mathbf{v} ~ v_x F^\text{L}_+(\mathbf{K},\bar{\omega},\mathbf{v}) \cos(qz), \\
J_{L,z}(z) &=& i \int \mathcal{D}q\mathcal{D}^3\mathbf{v} ~ v_z F^\text{L}_-(\mathbf{K},\bar{\omega},\mathbf{v}) \sin(qz). \label{jlz}
\end{eqnarray}
As we have discussed in Sec.~\ref{section:2}, the Drude current makes no net contributions. Moreover, as $J_{L,z}(0)\equiv 0$ from Eq.~(\ref{jlz}), we conclude that $\mathcal{P}_{s,b} = 0$. We then obtain
\begin{equation}
\mathcal{P}_{,b} = \mathcal{P}_{L,b} := \int dz ~ \left[J'_{L,z}(z)E_z(z) - J''_{L,x}(z)E_x(z)\right].
\end{equation}
On using $E_x(z)\approx E_z(z) \approx 2\pi\rho_s e^{-kz}$ outside the layer of surface charges, this expression can be brought into the following form,
\begin{equation}
\mathcal{P}_{L,b} = - 2\pi\rho_s \int \tilde{\mathcal{D}}q\mathcal{D}^3\mathbf{v} ~ \frac{\mathbf{K}\cdot\mathbf{v}}{k^2+q^2}\left(\frac{\mathbf{K}\cdot\mathbf{v}}{\bar{\omega}}\frac{\mathbf{K}\cdot\mathbf{v}}{\bar{\omega}-\mathbf{K}\cdot\mathbf{v}}\right)''. \label{31}
\end{equation}
Here we have introduced another shorthand $$\int \tilde{\mathcal{D}}q\mathcal{D}^3\mathbf{v} ... = \left(\frac{m}{2\pi\hbar}\right)^3\int^{\infty}_{-\infty}dq \frac{4\rho_q}{k^2+q^2}\int d^3\mathbf{v}\left(-e^2f'_0\right) ... $$ In this shorthand, $\rho_{q<0}:= \rho_{-q}$ has been implicitly understood. 

Note that $\mathcal{P}_{L,b}$ is always positive and it brings about damping even in the limit $\gamma_0\rightarrow 0_+$. This is so because the integrand has a pole located at $\omega_s = \mathbf{K}\cdot\mathbf{v}$. For infinitesimal positive $\gamma_0$, one may take $\left(\frac{1}{\bar{\omega}-\mathbf{K}\cdot \mathbf{v}}\right)'' \approx - \pi \delta(\omega_s - \mathbf{K}\cdot \mathbf{v})$. In this limit, the damping corresponds to nothing but the usual Landau damping~\cite{jellium}. As is well known, the rate of Landau damping incurred by SPWs is of the order of $kv_F$. This statement is easily confirmed with Eq.~(\ref{31}). Performing the integration with $\rho_q \approx \rho_s$ yields for infinitesimal positive $\gamma_0$ the following
\begin{equation}
\mathcal{P}_{L,b}/\mathcal{E}_p \approx \frac{3}{2}\frac{\omega^2_p}{\omega^2_s}kv_F. \label{32}
\end{equation}
Had we ignored $\mathbf{J}_s$, we would reach by means of this equation and Eq.~(\ref{key}) the well known result for SPW damping in the hydrodynamic-Drude model with inclusion of the Landau damping, i.e. $$\gamma_\text{HD} = -\frac{1}{\tau}- \frac{3}{4}\frac{\omega^2_p}{\omega^2_s}~kv_F.$$ This formula has often been used to estimate the electronic collision rate $\tau^{-1}$ by measurement of the line width of electron energy loss spectra (EELS) due to the excitation of SPWs. As to be seen in what follows, including the contribution of $\mathbf{J}_s$ calls into question the validity of this procedure. 

We should point out that, the as-established Landau damping would become a Landau gain if we assumed an infinitesimal negative $\gamma_0$. This is, of course, in violation of the causality principle and unphysical~\cite{landau1946}. Also in view of this, we must have $\gamma_0$ positive always. 

\section{Electrical work via $\mathbf{J}_s$}
\label{section:6}
Now we consider the contribution of $\mathbf{J}_s(z)$ to $\mathcal{P}_{b/s}$ and show that it would result in an instability of the system if not for thermal electronic collisions. Let us call the contribution $\mathcal{P}_{,s} = \mathcal{P}_{b,s}+\mathcal{P}_{s,s}$, where $\mathcal{P}_{s,s} = J'_{s,z}(0) \phi(0)$ and 
\begin{eqnarray}
\mathcal{P}_{b,s} = \int dz ~ \left[J'_{s,z}(z)E_z(z) - J''_{s,x}(z)E_x(z)\right],
\end{eqnarray}
Again using $E_x(z)\approx E_z(z) \approx 2\pi\rho_s e^{-kz}$ for $z$ outside the layer of surface charges, this expression can be rewritten as
\begin{equation}
\mathcal{P}_{b,s} = \Xi'_z - \Xi''_x, \quad \mathbf{\Xi} = 2\pi\rho_s \int dz ~ \mathbf{J}_s(z)e^{-kz}. 
\end{equation}
Using Eq.~(\ref{gs}), we find 
\begin{eqnarray}
\mathbf{\Xi} &=& 2\pi\rho_s\int \mathcal{D}q\mathcal{D}^3\mathbf{v} \Theta(v_z)\frac{iv_z\mathbf{v}}{\bar{\omega}-\mathbf{k}^*\cdot\mathbf{v}_-}\times \nonumber \\&~&\quad \quad \left[F_0(\mathbf{k},\bar{\omega},\mathbf{v}) - pF_0(\mathbf{k},\bar{\omega},\mathbf{v}_-)+2(p-1)F_+\right]. 
\end{eqnarray}
For small $kv_F/\omega_p$ and assuming $\rho_q \approx \rho_s$, it follows that
\begin{eqnarray}
&~&\mathcal{P}_{b,s} = -\frac{3+p}{2}\frac{\pi \rho^2_s}{k}\frac{3kv_F}{4}\frac{\omega^2_p}{\omega^2_s} + (1-p)2\pi\rho_s \times \nonumber \\ &~& \quad \quad \quad \int \mathcal{D}q\mathcal{D}^3\mathbf{v}\Theta(v_z)v_z\left[v_x\left(\frac{2F_+}{\bar{\omega}}\right)'+v_z\left(\frac{2F_+}{\bar{\omega}}\right)''\right].
\end{eqnarray}
By virtue of the separation $F_+ = F^D_+ + F^L_+$ as defined in Eq.~(\ref{fl}), we obtain
\begin{eqnarray}
&~& \mathcal{P}_{b,s} = \mathcal{P}_{L,s} - \frac{1+3p}{2}\frac{\pi \rho^2_s}{k}\frac{3kv_F}{4}\frac{\omega^2_p}{\omega^2_s} + (1-p)2\pi\rho_s \times \nonumber \\ 
&~ & \quad \quad \quad \quad \int \tilde{\mathcal{D}}q\mathcal{D}^3\mathbf{v}\Theta(v_z)\left(\frac{v_zv_x}{\bar{\omega}}\frac{\mathbf{K}\cdot\mathbf{v}}{\bar{\omega}}\frac{\mathbf{K}\cdot\mathbf{v}}{\bar{\omega}-\mathbf{K}\cdot \mathbf{v}}\right)', \label{37}
\end{eqnarray}
where 
\begin{eqnarray}
\mathcal{P}_{L,s} &=& (1-p)2\pi\rho_s \times \nonumber \\ &~& \quad \quad \int \tilde{\mathcal{D}}q\mathcal{D}^3\mathbf{v}\Theta(v_z)\left(\frac{v^2_z}{\bar{\omega}}\frac{\mathbf{K}\cdot\mathbf{v}}{\bar{\omega}}\frac{\mathbf{K}\cdot\mathbf{v}}{\bar{\omega}-\mathbf{K}\cdot \mathbf{v}}\right)'', \label{38}
\end{eqnarray}
which has a similar form as Eq.~(\ref{31}). One might think that this should give the Landau damping stemming from $\mathbf{J}_s(z)$. However, instead of damping, it actually represents a gain. In the same manner as we evaluated $\mathcal{P}_{L,b}$ assuming infinitesimal positive $\gamma_0$, we find
\begin{equation}
\mathcal{P}_{L,s}/\mathcal{E}_p \approx \frac{3(p-1)}{4} \frac{\omega^2_p}{\omega^2_s}~kv_F,  
\end{equation}
which is negative, i.e. it counteracts Landau damping rather than reinforces it. The last term in Eq.~(\ref{37}) contributes a higher order term in $kv_F/\omega_p$. Because of the presence of $v_x$ in the integrand, the contribution of that term after divided by $\mathcal{E}_p$ goes like $(kv_F/\omega_p)^2$ and may be neglected in the first order approximation. As such, we establish that
\begin{equation}
\mathcal{P}_{b,s}/\mathcal{E}_p \approx -\frac{3(3+p)}{8} \frac{\omega^2_p}{\omega^2_s} ~ kv_F. \label{40}
\end{equation}
It should be pointed out that this term alone already wins over Landau damping, as $\mathcal{P}_{L,b} + \mathcal{P}_{b,s}\leq0$ always [c.f. Eq.~(\ref{32})]. The inclusion of $\mathcal{P}_{b,s}$ in the present calculation supplements the calculation reported in Ref.~\cite{deng2016a,deng2016b} in the following technical matter. In previous work~\cite{deng2016a,deng2016b}, we omitted the there-named $\mathcal{M}$ matrix; see Appendix B in Ref.~\cite{deng2016a}. The effects of this matrix in the equation of motion for the charge density are tantamount to those of $\mathcal{P}_{b,s}$ in energy conversion. 

The calculation of $\mathcal{P}_{s,s}$ can be performed in a straightforward manner. We obtain
\begin{equation}
\mathcal{P}_{s,s}/\mathcal{E}_p =  - 2\Gamma(\gamma_0) + 2b\gamma_0, \quad b = \frac{1+p}{4}\frac{\omega^2_p}{\omega^2_s}<1, \label{41}
\end{equation} 
where $\Gamma(\gamma_0)$ is a function of $\gamma_0$ given by
\begin{equation}
\Gamma(\gamma_0) = (1-p)\rho^{-1}_s\int \tilde{\mathcal{D}}q\mathcal{D}^3\mathbf{v}\Theta(v_z)v_z\left(\frac{\mathbf{K}\cdot\mathbf{v}}{\bar{\omega}}\frac{\mathbf{K}\cdot\mathbf{v}}{\bar{\omega} - \mathbf{K}\cdot\mathbf{v}}\right)'.  
\end{equation}
Note that $\Gamma$ is positive and it dominates all the contributions from other parts of $\mathcal{P}_{b/s}$. Actually, we have
\begin{equation}
2\gamma_L = - (\mathcal{P}_{L,b}+\mathcal{P}_{b,s})/\mathcal{E}_p \sim (kv_F/\omega_p) ~ \Gamma. 
\end{equation}
Combining Eqs.~(\ref{32}), (\ref{40}) and (\ref{41}), we can recast the energy balance equation~(\ref{key}) as follows
\begin{equation}
\Gamma(\gamma_0) + \gamma_L - (1+b)\gamma_0 = 0.
\label{key2}
\end{equation}
We note that $\gamma_L$ may be regarded as the net contribution to $\gamma_0$ from the usual electrical work, i.e. $\mathcal{P}_b$. 

If we use Eqs.~(\ref{32}) and (\ref{40}) as an estimate of $\mathcal{P}_{L,b}$ and $\mathcal{P}_{b,s}$ respectively, we obtain 
\begin{equation}
\quad \gamma_L \approx \frac{\omega^2_p}{\omega^2_s}\frac{3(1-p)}{16}kv_F, \label{45}
\end{equation}
which is always non-negative. From this we may conclude that even if $\mathcal{P}_{s,s}$ is totally ignored, the Landau damping as caused by $\mathbf{J}_b(z)$ will still be overcompensated by the gain due to $\mathbf{J}_s(z)$. As such, in contrast to what is claimed by J. Khurgin \textit{et al}. and in agreement with observations, Landau damping does not constitute an unsurmountable loss barrier in sub-wavelength plasmonics. The long-standing puzzle, as highlighted in the book by Raether~\cite{raether1988}, of the apparent weak coupling of SPWs to single particle excitations -- the origin of Landau damping -- becomes thus explicable: such coupling is not weak but just overshadowed by surface effects. 

In the next section, we shall solve Eq.~(\ref{key2}) and show that $\gamma_0>0$ invariably, in agreement with what is expected of the causality principle [c.f. the remark below Eq.~(\ref{gs})]. By Eq.~(\ref{45}), one might wrongly think that $\gamma_0$ should vanish for $p=1$. This is not true, because Eq.~(\ref{45}) is based on the estimate of $\mathcal{P}_{L,b}$ by Eq.~(\ref{32}), which assumes an infinitesimal $\gamma_0$. For actual $\gamma_0$, equation (\ref{32}) only gives an overestimate of $\mathcal{P}_{L,b}$. Thus, $\gamma_L$ is always positive even for specularly reflecting surfaces. Now that the net damping rate is $$-\gamma = -\omega'' =  1/\tau - \gamma_0,$$ the positiveness of $\gamma_0$ then implies that the system, or more precisely the Fermi sea, is unstable if $\tau$ is sufficiently large. In the meanwhile, if we approach the instability point from the side where the system is stable, we can in principle make $ - \gamma$ as small as required provided that $1/\tau$ can be tuned below $\gamma_0$. This observation calls into question the practice of identifying $\tau^{-1}$ with the line width of the peak due to excitation of SPWs in for example EELS, which is $-\gamma$. 

\section{Instability and Amplification}
\label{section:7}
We have demonstrated that the electrical current in a semi-infinite metal generally consists of two components, which we call the bulk and the surface components, respectively. The bulk component describes bulk electrical responses of the system and naturally extends the hydrodynamic theory to account for Landau damping. The surface component, however, arises only in the presence of a surface and hence describes purely surface effects. It is totally beyond the scope of the hydrodynamic model. The bulk and surface components play disparate roles in energy conversion. While the bulk component would basically preserve the kinetic energy of the electrons if not for Landau damping, the surface component transfers it to the SPWs. This picture of energy conversion is summarized in Eq.~(\ref{key2}), which we now solve to demonstrate that an instability of the Fermi sea might take place under certain circumstances.  

Before we numerically solve Eq.~(\ref{key2}), let us note that although the formalism derived of the energy conversion in this work is exact, equation (\ref{key2}) does not provide a complete picture of SPWs on its own. The reason is because it does not afford a means of evaluating $\omega_s$ and $\rho_q$ at the same time. In the complete description established in Ref.~\cite{deng2016a}, these quantities were determined self-consistently. In the present work, we have provided them in a reasonable yet \textit{ad hoc} manner. These provisions are consistent with the complete description and the details of them are not important in the energy conversion process in question here. 

We obtain $\gamma_0$ by first numerically evaluating both $\gamma_L$ and $\Gamma$ as functions of $\gamma_0$ and then inserting them in Eq.~(\ref{key2}), which is further solved outright. The results are displayed in Fig.~\ref{figure:f2}. In panel (a), the $k$ dependence is shown at fixed $p$, where we see that $\gamma_0$ linearly decreases as $k$ increases. Roughly, $\gamma_0(p,k)/\omega_p \approx 0.16 - 0.25k/k_p$ for $p=0$, in good agreement with what was found in our previous work~\cite{deng2016}. It should be noted that this relation is universal in the sense that it is regardless of the material parameters, which -- in the jellium model -- are signified by $\omega_p$ and $k_p$ only. In panel (b), we show the $p$ dependence of $\gamma_0$ at fixed $k$. Again a linear dependence develops here, $\gamma_0(p,k)/\omega_p \approx 0.12 - 0.066 p$ for $k=0.07k_p$. Combined, we may fit the numerical solutions by the following function $$\gamma_0(p,k)/\omega_p \approx 0.16 - 0.25k/k_p - 0.066p.$$ As explained in the remarks made in the last paragraph of the preceding section, $\gamma_0$ remains finite even for $p=1$. See that the $p$ dependence disagrees with what was found in previous work. This discrepancy occurs because in this work we have used a different boundary condition for the electronic distribution function [c.f. Eq.~(\ref{bc})]. In Ref.~\cite{deng2016a}, the corresponding condition assumes $p$ as the probability of specular reflection only in the absence of the normal component of the electric field. In the present work, $p$ is the probability for any electric field and thus more realistic. Nevertheless, these two conditions are identical for diffuse boundaries.  

In closing this section, we discuss some experimental evidences and repercussions for the results. In the preceding section, we have noted that the apparent experimental absence of the coupling between SPWs and single particle excitations, which is supposed to give rise to pronounced Landau damping, is well explicable in our theory; see the remarks following Eq.~(\ref{45}). Apart from this, other evidence in support of the existence of the intrinsic channel of gain may be found by comparing these two rates: the directly measured relaxation rate $1/\tau$ at SPW frequencies and the directly measured SPW damping rate $-\gamma$. We expect a substantial difference between them if the intrinsic channel is not suppressed. For materials in which inter-band transitions can be neglected at both the frequencies of SPWs and bulk plasma waves, we may take the  bulk wave damping rate as a measure of $1/\tau$. In such case, the SPW damping rate should be substantially less than the bulk wave damping rate. At least for two alkali metals, potassium (K) and cesium (Ce), this proposition is confirmed~\cite{beck1991}: the damping rates for SPWs and bulk waves in K are 0.1eV and 0.24eV, respectively, while those in Ce are 0.23eV and 0.75eV, respectively. However, the situation with other materials remains unclear. 

Several factors may contribute to suppress the intrinsic channel. Firstly, there is a size effect. In Ref.~\cite{deng2016b}, we showed that for metal films $\gamma_0$ decreases quickly to zero as the film thickness decreases below the SPW wavelength. Secondly, inter-band transitions and extra losses due to surface scattering and radiation can also reduce the value of $\gamma_0$; see discussions in the next section. In addition, if the metal is in contact with a dielectric rather than the vacuum, $\gamma_0$ may also be affected. Some of these effects have recently been addressed in Ref.~\cite{deng2017} and the intrinsic channel of gain survives. A detailed discussion of them is beyond the scope of the present paper. 

\section{Summary}
\label{section:8}
On the basis of the continuity equation and the Boltzmann theory, we have presented a systematic analysis of the energy conversion in SPWs supported in semi-infinite metals. An important role played by the surface is revealed in the conversion process. We find that ballistic motions could destabilize the system and lend SPWs an intrinsic amplification channel with a rate $\gamma_{0}$ given as the solution to Eq.~(\ref{key2}). Via this channel, SPWs can extract energy from the Fermi sea and amplify themselves if the loss channels are sufficiently suppressed. The positiveness of $\gamma_0$ is actually warranted by the principle of causality. 

In the present work we have explicitly considered the losses due to Landau damping and thermal electronic collisions. The Landau damping has been shown to be a higher order effect in comparison with the intrinsic gain and negligible at long wavelengths, while electronic collisions directly counteracts the gain. For any real materials, of course there are additional losses such as inter-band absorption and radiative losses. We briefly discuss these losses in what follows. 

Inter-band transitions not only give rise to losses but also modify the SPW frequency $\omega_s$. A systematic treatment of these effects can only be achieved by self-consistently solving the basic equation of motion for the charge density, as we did in Ref.~\cite{deng2016a}. In regard to energy conversion, however, one may obtain a qualitative appreciation by resorting to a simple picture. Inter-band effects stem from the electrical responses of inner-shell electrons of the atoms in the metal, namely, the valence electrons. These electrons are usually tightly bound to their host atoms and less susceptible to the presence of atomic surroundings and sample boundaries. As such, one may reasonably assume a spatially non-dispersive response function to describe the motions of these electrons under an an electric field. The current from such motions can then be written as $\mathbf{J}_p(z) = \sigma_p(\omega)\mathbf{E}(z)$. Note that the conductivity $\sigma_p(\omega)$ can be related to a dielectric function $\epsilon_p(\omega) = 4\pi i\sigma_p(\omega)/\omega$, which can be determined experimentally or \textit{ab initio}. Then the rate of inter-band absorption can be estimated as $(1/2\mathcal{E}_p)\int^\infty_0 dz ~ \sigma'_p(\omega) E^2(z) \approx \pi \sigma'_p(\omega_s)$. Depending on whether $\omega_s$ stays close to the inter-band transition threshold or not, this rate can be significant or negligible. Despite this uncertainty, in a future paper we will show that inter-band absorption, though capable of diminishing the intrinsic gain, can not erase it in total. 

Radiative losses occur due to the transmutation of a plasmon into a propagating photon. In the case of an ideally flat metal surface, this process can not happen because of the conservation of both energy and momentum. With a non-flat surface, the momentum conservation is lifted and the transmutation takes place in the form of optical scattering. By means of a dimensional analysis and some simple physical arguments~\cite{raether1988}, the associated loss rate may be determined as $\sim \omega_s (k^2_0\sigma \delta)^2$, where $k_0$ is the wavenumber of emitted light of frequency $\omega_s$ while $\sigma$ and $\delta$ are the mean squared height fluctuation and the variation, which characterize the profile of a Gaussian surface. This expression also gives an estimate of the losses due to SPW scattering by surface roughness. These losses simply add to $1/\tau$. Considering that $k^{-1}_0 \sim 100$nm and $\sigma \sim \delta \sim 0.1$nm typically, we may safely ignore them in most cases. 

In summary, we have derived a generic formalism for studying energy conversion processes in systems with boundaries. We have applied it to SPWs and shown that the Fermi sea of metals is unstable thanks to the interplay between ballistic electronic motions and the presence of a surface. 

We hope this work will stimulate more interest in this new aspect of SPWs from both the theoretical and experimental communities. 
\\
\\
\textbf{Acknowledgement} --
This work had commenced when the author was affiliated with Kwansei Gakuin University, Japan, where he was a JSPS Research Fellow supported by the International Research Fellowship of the Japan Society for the Promotion of Science (JSPS). 

\appendix*
\section{The surface term in Eq.~(\ref{1})}
\label{sec:b}
The continuity equation, Eq.~(\ref{1}) may also be understood from Boltzmann's equation in the relaxation time approximation, which is written as $$\left(\partial_t+\tau^{-1}+\mathbf{v}\cdot\partial_{\mathbf{x}}\right)g(\mathbf{x},\mathbf{v},t)+\frac{\mathbf{F}}{m}\cdot\partial_{\mathbf{v}}f_0(v) = -\frac{\mathbf{F}}{m}\cdot \partial_{\mathbf{v}}g(\mathbf{x},\mathbf{v},t),$$ where $f_0$ and $g$ are the equilibrium and non-equilibrium part of the Boltzmann distribution function, respectively, $m$ is the electron mass and $\mathbf{F}$ is the total force (excluding the part taken care of by the $\tau^{-1}$ term) acting on the electrons. We may write $$\mathbf{F}=e\mathbf{E}+\mathbf{F}_s,$$ where $e$ is the electron charge and $\mathbf{F}_s$ is the force that prevents the electrons from escaping the metal. We then write $$\mathbf{F}_s=e\mathbf{E}_s=-e\partial_{\mathbf{x}}\phi_s,$$ with $\phi_s$ denotes the surface potential. For an ideal surface $\mathbf{F}_s$ should vanish everywhere except on the surface and point normal to the surface. Keeping $g$ to the first order in $\mathbf{E}$, we can write $$(\mathbf{F}/m)\cdot\partial_{\mathbf{v}}g=(\mathbf{F}_s/m)\cdot\partial_{\mathbf{v}}g.$$ Now multiplying the equation by $e (m/2\pi\hbar)^3$ and integrating it over $\mathbf{v}$, we find $$(\partial_t+\tau^{-1})\rho+\partial_{\mathbf{x}}\cdot\mathbf{J}=-e(m/2\pi\hbar)^3\int d\mathbf{v}(\mathbf{F}_s/m)\cdot \partial_{\mathbf{v}}g.$$ We cannot proceed further without knowing $\mathbf{F}_s$, whose details are generally difficult to know and could vary greatly from one sample to another. Despite this, we can fix it by demanding that the equation of continuity holds, i.e. $$e(m/2\pi\hbar)^3\int d\mathbf{v}(\mathbf{F}_s/m)\cdot \partial_{\mathbf{v}}g = \delta(z)J_z(\mathbf{x},t).$$ This then leads to Eq.~(\ref{1}). This derivation makes it clear that this term can be traced back to the force exerted by the surface. From Boltzmann's equation, one can show that the total energy $E_k+E_s+E_p$ is conserved for $\tau\rightarrow\infty$, where $$E_k(t)=(m/2\pi\hbar)^3\int d\mathbf{x}\int d\mathbf{v}(m/2)\mathbf{v}^2\left(f_0(\mathbf{v})+g(\mathbf{x},\mathbf{v},t)\right)$$ is the kinetic energy and $E_s(t) = \int d\mathbf{x}\phi_s(\mathbf{x})\rho(\mathbf{x},t)$ as well as $E_p(t)= (1/2)\int  d\mathbf{x}\phi(\mathbf{x},t)\rho(\mathbf{x},t).$


\begin{thebibliography}{9}
\bibitem{note2}
If we take $\hbar\omega_s \approx \sqrt{\frac{2\pi ne^2\hbar}{m}}$ and $\varepsilon_F=\frac{\hbar^2k^2_F}{2m}$ as well as $k^3_F = 6\pi^2n$, we obtain $\frac{\hbar\omega_s}{\varepsilon_F} = \frac{\sqrt{8\pi}}{(6\pi^2)^{2/3}}\left(\frac{n_0}{n}\right)^{1/6}\approx 0.5\left(\frac{n_0}{n}\right)^{1/6}$, where $n_0=\left(\frac{e\sqrt{m}}{\hbar}\right)^6\approx 10^{24}$cm$^{-3}$ comparable to $n$ in magnitude. Thus, $\hbar\omega_s \sim \varepsilon_F$ and $k_s \sim k_F \sim n^{1/3}$. As such, we have $q_c\sim k_s$. In numerical calculations, we use $q_c=1.5k_p$ in this paper as in previous work. In Ref.~\cite{deng2017}, we have demonstrated that the values of $\omega_s+i\gamma_0$ as solutions to the fundamental equation of SPWs [c.f. Eq.~(18) in Ref.~\cite{deng2016a}] are independent of the choice of $q_c$ as long as the latter is sufficiently large -- convergence is quickly achieved for $q_c$ beyond $\sim k_p$.
\end{thebibliography}

\end{document}